\documentclass{sola}

\usepackage{natbib}
\usepackage{graphicx}
\usepackage{amsmath}
\usepackage{bm}
\renewcommand{\vec}[1]{\bm{#1}}

\def\affiliation#1#2{\centerline{\footnotesize \it $^{#1}$#2}}

\begin{document}


%

\title{Convex Optimization of Initial Perturbations toward Quantitative Weather Control}

%

\author{Toshiyuki Ohtsuka\affil{1}, Atsushi Okazaki\affil{2,3}, Masaki Ogura\affil{4}, and Shunji Kotsuki\affil{2,3}}

\affiliation{1}{Graduate School of Informatics, Kyoto University, Kyoto, Japan}
\affiliation{2}{Institute for Advanced Academic Research, Chiba University, Chiba, Japan}
\affiliation{3}{Center for Environmental Remote Sensing, Chiba University, Chiba, Japan}
\affiliation{4}{Graduate School of Advanced Science and Engineering, Hiroshima University, Hiroshima, Japan}

\correspondingauthor{Toshiyuki Ohtsuka}{Graduate School of Informatics, Kyoto University}{Yoshida-honmachi, Kyoto 606-8501, Japan}{ohtsuka@i.kyoto-u.ac.jp}

%

\runningtitle{Convex Optimization of Initial Perturbations toward Quantitative Weather Control}
\runningauthor{Ohtsuka et al.}


%

\begin{abstract}
This study proposes introducing convex optimization to find initial perturbations of atmospheric states to realize specified changes in subsequent weather. 
In the proposed method, we formulate and solve an inverse problem to find effective perturbations in atmospheric variables so that controlled variables satisfy specified changes at a specified time. 
The proposed method first constructs a sensitivity matrix of controlled variables, such as accumulated precipitation, to the initial atmospheric variables, such as temperature and humidity, through sensitivity analysis using a numerical weather prediction (NWP) model. 
Then a convex optimization problem is formulated to achieve various control specifications involving not only quadratic functions but also absolute values and maximum values of the controlled variables and initial atmospheric variables in the cost function and constraints. 
The proposed method was validated through a benchmark warm bubble experiment using the NWP model.
The experiments showed that the identified perturbations successfully realized specified spatial distributions of accumulated precipitation. 
\end{abstract}

%

\section{Introduction}

Climate change is expected to intensify severe weather-induced disasters in the future. For example, tropical cyclones, which often cause significant social impacts, are predicted to intensify, along with the corresponding heavy rainfall \citep{IPCC_AR6_WG1}. Mitigating the impacts of such extreme weather events is an urgent task. Weather modification, which intentionally manipulates or alters the atmosphere, can be an option as an adaptation countermeasure here. The feasibility of weather modification has long been explored \citep[see][and references therein]{Bruintjes99}, mainly to increase precipitation by cloud seeding. 
However, no study has succeeded in controlling real-world severe weather by weather modifications. 
In this context, countermeasures to mitigate weather-induced disasters are explored in a Japanese research program called the Moonshot Program, in which the mitigation of severe weather with feasible interventions is considered a major challenge \citep{Spectrum23}. 

Recent studies have started exploring a method to utilize the chaotic nature of the atmosphere to mitigate severe weather with small intervention \citep{MS22,SMR23,OTK23,KK24}. In their groundbreaking studies, \citet{MS22} proposed a Control Simulation Experiment (CSE) framework to control the chaotic system, where small perturbations are added to \textit{nature} to guide the system to a desirable trajectory. The feasibility of CSE has been demonstrated with the Lorenz-63 three-variable system \citep{Lorenz63} in \citet{MS22}, \citet{OTK23} and \citet{KK24}, and with the Lorenz-96 40-variable system \citep{LE98} in \citet{SMR23}, which further expanded the possibility of controlling the weather. 

However, there are still several challenges to overcome in the realization of quantitative weather control. 
One major challenge is engineering to develop practical intervention techniques to add perturbations to the atmosphere.
Furthermore, since such perturbations are considered to be very small compared to the energy of the atmosphere, these interventional perturbations need to be optimized to guide the atmospheric states effectively, which is the second major challenge. 
Specifically, in optimization of perturbations, there are three issues that need to be addressed. 
First, weather phenomena pose challenges for the application of standard control design methodologies from control engineering \citep[e.g.,][]{ConHandbook10,EncSysCon21}, due to the scale and complexity of realistic numerical weather prediction (NWP) models. 
Second, weather phenomena exhibit chaotic and nonlinear behaviors, possibly preventing accurate prediction and control over long time intervals. 
Third, various quantitative specifications must be handled in the control design, such as minimizing or imposing an upper limit on the maximum values of precipitation and interventions. 

This paper addresses the above optimization issues with a suitable problem setting. Here, we propose a convex optimization approach to compute the control inputs for weather control. In our approach, we consider perturbations in the initial conditions of atmospheric models, such as temperature and humidity, as control inputs for weather control without assuming a particular intervention means. 
Similar settings of determining state perturbations rather than relying on particular methods of intervention are commonly seen in the literature on weather control \citep[e.g.,][]{HHLNG05, HHLGN06} and useful for gaining insight into the requirements for intervention techniques. 

\begin{figure}
    \centering
    \includegraphics[width=12cm]{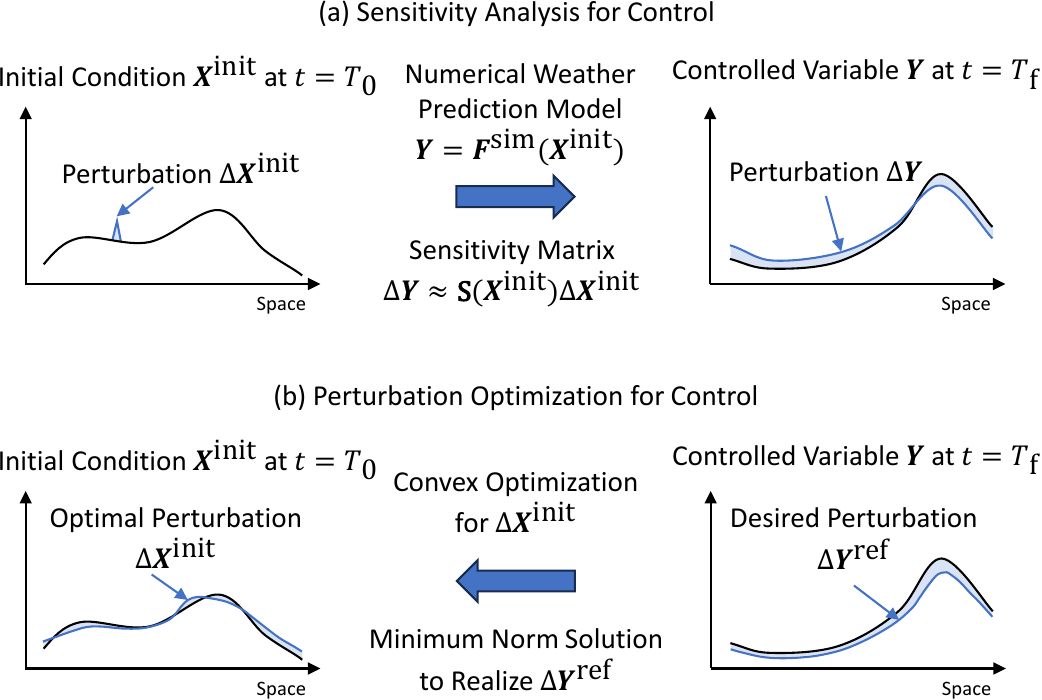} 
    \caption{(a) Sensitivity analysis for control and (b) convex optimization of perturbation in initial condition for realizing desired perturbation in controlled variable. }
    \label{fig:Schematic}
\end{figure}

In the proposed method, we first utilize an NWP model to construct a sensitivity matrix numerically between the perturbations in its initial conditions and targeting variables to be controlled at a specified final time, such as accumulated precipitation (Fig.\ \ref{fig:Schematic}a). 
This means that, in the construction of the sensitivity matrix, we do not need to manipulate the large-scale and complex equations of the NWP model symbolically; alternatively, we can construct a linear model numerically for predicting perturbations in the controlled variables. 
This approach is valid only for a sufficiently short time interval and sufficiently small perturbations where chaotic and nonlinear behaviors are not dominant.  
However, assuming small rather than drastic perturbations in atmospheric states and precipitation is generally reasonable because large perturbations are considered extremely difficult to realize in weather phenomena. 

We then introduce a convex optimization problem to find optimal perturbations in the initial conditions. This method allows us to realize the desired spatial distribution of the controlled variables (Fig.\ \ref{fig:Schematic}b), satisfying some control specifications given as constraints. 
Unlike existing methods for determining perturbations such as singular vectors \citep[e.g.,][]{DL12}, conditional nonlinear optimal perturbations \citep{MDWZ10}, and variational assimilation \citep[e.g.,][]{RJKMS00}, convex optimization provides greater flexibility in handling a wide range of cost functions and constraints. 
That is, convex optimization can deal with various cost functions and constraints, including not only quadratic functions, but also absolute values, maximum values, and general $\ell_p$ norms ($1 \leq p \leq \infty$) \citep[e.g.,][]{Boyd04}. 
In particular, a suitable choice of the cost function results in a sparse solution with many zero components, which is potentially advantageous in implementing weather control with a small number of local interventions. 
Moreover, a convex optimization problem is guaranteed to have a unique global optimal solution and can be solved efficiently. 
Therefore, convex optimization is particularly suitable for evaluating the achievable performance in weather control under various control specifications. 

By implementing our proposed method in a benchmark warm bubble experiment on a highly nonlinear NWP model, we demonstrate its ability to realize desired spatial distributions of accumulated precipitation, such as a reference distribution and the reduced maximum value. 
These results show the possibility of controlling a real weather phenomenon by perturbing atmospheric states as determined by the proposed method. 
Therefore, the proposed method can be a cornerstone in achieving quantitative weather control.

\section{Methods}
\subsection{Numerical weather prediction model and sensitivity analysis for control}\label{subsec:sensitivity}
An NWP model outputs some quantities, like accumulated precipitation, at the end of the simulation for given initial conditions of such atmospheric states as temperature, wind speed, and humidity over a region of interest. 
In this study, we consider realizing specified changes in some of the outputs of an NWP model as controlled variables by perturbing some of the initial conditions as manipulated variables (control inputs), which is formulated as an inverse problem.  
We denote a vector of the NWP model's controlled variables as $\vec{Y}$ and a vector of manipulated initial conditions in the region of interest as $\vec{X}^\mathrm{init}$. The superscript $\mathrm{init}$ denotes the initial condition. 
Then, the NWP model defines a function $\vec{F}^\mathrm{sim}$ as 
\begin{equation}
    \vec{Y} = \vec{F}^\mathrm{sim}(\vec{X}^\mathrm{init}), 
\end{equation}
where the superscript $\mathrm{sim}$ denotes the simulation.
If we assume smoothness of the function $\vec{F}^\mathrm{sim}$, we can consider its sensitivity matrix $\mathbf{S}(\vec{X}^\mathrm{init}) = \partial \vec{F}^\mathrm{sim}(\vec{X}^\mathrm{init}) / \partial \vec{X}^\mathrm{init}$ to predict the perturbation $\Delta \vec{Y}$ in the controlled variable $\vec{Y}$ for a perturbation $\Delta \vec{X}^\mathrm{init}$ in the initial condition $\vec{X}^\mathrm{init}$ as: 
\begin{equation}
    \Delta \vec{Y} \approx \mathbf{S}(\vec{X}^\mathrm{init}) \Delta \vec{X}^\mathrm{init}.
\end{equation} 
Since an explicit expression of $\vec{F}^\mathrm{sim}$ is unavailable or too complex in practice, we construct the $(i,j)$ element $\mathbf{S}_{ij}(\vec{X}^\mathrm{init})$ of the sensitivity matrix by finite difference with a small step $h$, for example, as 
\begin{equation}
    \mathbf{S}_{ij}(\vec{X}^\mathrm{init}) \approx \frac{\vec{F}_i^\mathrm{sim}(\vec{X}^\mathrm{init} + h \vec{e}_j) - \vec{F}_i^\mathrm{sim}(\vec{X}^\mathrm{init})}{h} . 
\end{equation}
$\vec{F}_i^\mathrm{sim}$ denotes the $i$th component of $\vec{F}^\mathrm{sim}$, and $\vec{e}_j$ denotes a standard basis vector with only the $j$th component one and other components zero. 
We can choose other methods for numerical computation of the sensitivity matrix and any other basis vectors in general. 
Note that we define indices of $\vec{Y}$ and $\vec{X}^\mathrm{init}$ with appropriate orderings of the grids or basis vectors in the NWP model over their regions of interest, respectively. 

Although the computation cost of the sensitivity matrix over all grids is high for large-scale models, we can reduce the computation cost by reducing the number of optimized variables and exploring alternative computational methods or adopting different models to handle the sensitivity analysis more efficiently. 
We can reduce the number of optimized variables by identifying effective grids or suitable bases for perturbations, which is one of our future work. 
The second possibility is to reduce the computational cost for sensitivity analysis employing the adjoint method \citep{CR97} with the ensemble approximation or AI-based weather prediction models \citep[e.g.,][]{GraphCast23,FourCastNet22} that have developed rapidly in recent years. 
In any case, we can freely choose the grids or bases for perturbation in control synthesis and make a trade-off between the computation cost and achievable control performance. 
In this paper, we do not introduce any approximation or simplification, except for the finite difference approximation of the sensitivity matrix, to evaluate the best possible performance of convex optimization.

\subsection{Convex optimization of perturbation for control}
We propose a convex optimization method to determine a perturbation $\Delta \vec{X}^\mathrm{init}$ in the initial condition $\vec{X}^\mathrm{init}$ to realize a desired perturbation $\Delta \vec{Y}^\mathrm{ref}$ in the controlled variable $\vec{Y}$. 
We assume there are sufficient degrees of freedom for control. 
That is, the solution $\Delta \vec{X}^\mathrm{init}$ to a linear equation $\mathbf{S}(\vec{X}^\mathrm{init}) \Delta \vec{X}^\mathrm{init} = \Delta \vec{Y}^\mathrm{ref}$ always exists for any $\Delta \vec{Y}^\mathrm{ref}$ but may not be unique in general. 
To find an optimal perturbation of the initial condition, we formulate a constrained norm minimization problem: 
\begin{equation}
    \min_{\Delta \vec{X}^\mathrm{init}} \| \Delta \vec{X}^\mathrm{init} \| \quad \textrm{subject to} \quad \mathbf{S}(\vec{X}^\mathrm{init}) \Delta \vec{X}^\mathrm{init} = \Delta \vec{Y}^\mathrm{ref},  \label{eq:CMNP} 
\end{equation}
where $\| \Delta \vec{X}^\mathrm{init} \|$ is a vector norm. 
For example, $\ell_2$ norm 
\begin{equation}
    \| \Delta \vec{X}^\mathrm{init} \| = \sqrt{\sum_{j=1}^{N_X} \left( \vec{X}^\mathrm{init}_j \right)^2},   
\end{equation}
and $\ell_1$ norm
\begin{equation}
    \| \Delta \vec{X}^\mathrm{init} \| = \sum_{j=1}^{N_X} \left| \vec{X}^\mathrm{init}_j \right|,   
\end{equation}
are widely used. 
It is well known that the $\ell_1$ norm minimization results in a sparse solution, i.e., a vector with many zero components \citep[e.g.,][]{Boyd04}. 

The general form of the constrained norm minimization problem is given by: 
\begin{eqnarray}
    \min_{\Delta \vec{X}^\mathrm{init}} \| \Delta \vec{X}^\mathrm{init} \| 
    \quad 
    \textrm{subject to} \quad 
    \mathbf{A} \Delta \vec{X}^\mathrm{init} \leq \vec{b}, \quad \mathbf{A}_\mathrm{eq} \Delta \vec{X}^\mathrm{init} = \vec{b}_\mathrm{eq},  \label{eq:GeneralCMNP} 
\end{eqnarray}
where $\mathbf{A}$ and $\mathbf{A}_\mathrm{eq}$ are matrices, $\vec{b}$ and $\vec{b}_\mathrm{eq}$ are vectors of consistent dimensions, and the inequality is imposed component-wise. 
The constrained norm minimization problem is a convex optimization problem with a unique global optimal solution, for which various efficient numerical solvers are available \citep[e.g.,][]{Boyd04}. 

In general, a suitable norm can be chosen to be minimized according to the achievable control performance and the required interventions. 
Although minimization of any norm is formulated as a convex optimization problem and can be solved efficiently, 
the $\ell_2$ norm and the $\ell_1$ norm are most commonly used because minimization of these norms reduces to a quadratic programming problem (QP) and a linear programming problem (LP), respectively. 
QP and LP can be solved quite efficiently, and there are many reliable and easy-to-use solvers for QP and LP. 

The optimization problem (7) can represent a wide variety of problems, such as constraints on the magnitudes of manipulated initial conditions (control inputs) and controlled variables (outputs). 
In particular, we can also minimize or bound the maximum values of their magnitudes. 
This flexibility in control specifications is one of the advantages of the convex optimization approach.

\section{Experiments and results}
\subsection{Settings of experiments}
We validated the perturbation optimization for weather control in a warm bubble experiment, a widely used benchmark for weather forecasting studies \citep[e.g.,][]{ZS04}. 
We used an NWP model, SCALE-RM ver 5.4.5 \citep{Nishizawa15, Sato15}, for numerical experiments.\footnote{The source codes of SCALE-RM are available at https://scale.riken.jp/.} 
Numerous studies on weather forecasting use the regional non-hydrostatic model SCALE-RM \citep[e.g.,][]{Honda22}.
SCALE-RM prepares multiple schemes for each model component. 
This study used a 6-class single-moment bulk scheme \citep{Tomita08} for cloud microphysics. 
For simplicity, the model in this study does not consider turbulence or radiation. 
The model is configured to cover a two-dimensional domain of $y$-$z$ plain with a grid resolution of 500 m in the horizontal directions and 97 vertical layers with the model top of 20 km. In the warm bubble experiment, the initial vertical profile is given by \citet{Redelsperger00} with a wind profile of \citet{Oyama01}. A warm bubble with horizontal radius of 4 km and vertical radius of 3 km with a maximum intensity of 3 K is added to the center of the domain in a south-north direction.
Figure S\ref{supp:nominal} shows snapshots of the mass concentration of the total hydrometeor and precipitation intensity in the warm bubble experiment. 

We considered a vector of accumulated precipitation at the horizontal grids at the end of the simulation as the controlled variable $\vec{Y}$ of the NWP model. 
We constructed the sensitivity matrix $\mathbf{S}(\vec{X}^\mathrm{init})$ for either initial values of the product of density and potential temperature, $\rho \theta$, or initial values of specific humidity $q_v$ over all grid points.
Here we considered that manipulating $\rho \theta$ and $q_v$ would be easier than manipulating other atmospheric states such as wind and pressure fields. 
We chose the step size $h$ in the finite difference approximation as $h=0.1$ for $\rho \theta$ and $h=0.001$ for $q_v$, respectively. 
Then, we solved the constrained norm minimization problems to determine the perturbations in the initial values for realizing the desired distributions of accumulated precipitation at the end of the simulation. 
This study employed two types of desired distributions: reference values and a uniform upper bound over all grid points. 
Although we used the linear model of the perturbations in the constrained norm minimization problem, we applied the optimized perturbations to the highly nonlinear NWP model for validation. 

We used a PC with CPU Core i7-1165G7 2.8GHz, RAM 32GB, OS Windows 11 Pro, built SCALE-RM with gcc and gfortran on WSL2 (Ubuntu 22.04.3 LTS) and ran MATLAB scripts for sensitivity analysis and norm minimization problems on MATLAB 2023a.\footnote{The source codes for sensitivity analysis and perturbation optimization are available at https://
github.com/ohtsukalab/Weather-Control-by-InitCond.} 
We used the MATLAB lsqlin function for the $\ell_2$ norm minimization and the linprog function for the $\ell_1$ norm minimization \citep{OptimizationTB}. 
The execution time for one warm bubble experiment was 4 s, which amounts to 4 h for all 40$\times$97 perturbations. 
After obtaining history data for all perturbations, the computation time for constructing the sensitivity matrix by finite difference was 5 min.

\subsection{Realizing reference precipitation} \label{subsec:RefPrec}
First, we considered a control problem of reducing the accumulated precipitation to 90\% of the nominal value at each grid point. 
We set $\vec{Y}^\mathrm{ref} = 0.9 \vec{Y}$, i.e., $\Delta \vec{Y}^\mathrm{ref} = -0.1 \vec{Y}$, and solved the constrained norm minimization problem in Eq.\ (\ref{eq:CMNP}) for $\rho \theta$ with the $\ell_2$ norm and the $\ell_1$ norm, respectively. 
The computation times for solving the constrained norm minimization problems were 1 s and 6 s with the $\ell_2$ norm and $\ell_1$ norm, respectively. 

\begin{figure}
    \centering
    \includegraphics[width=12cm]{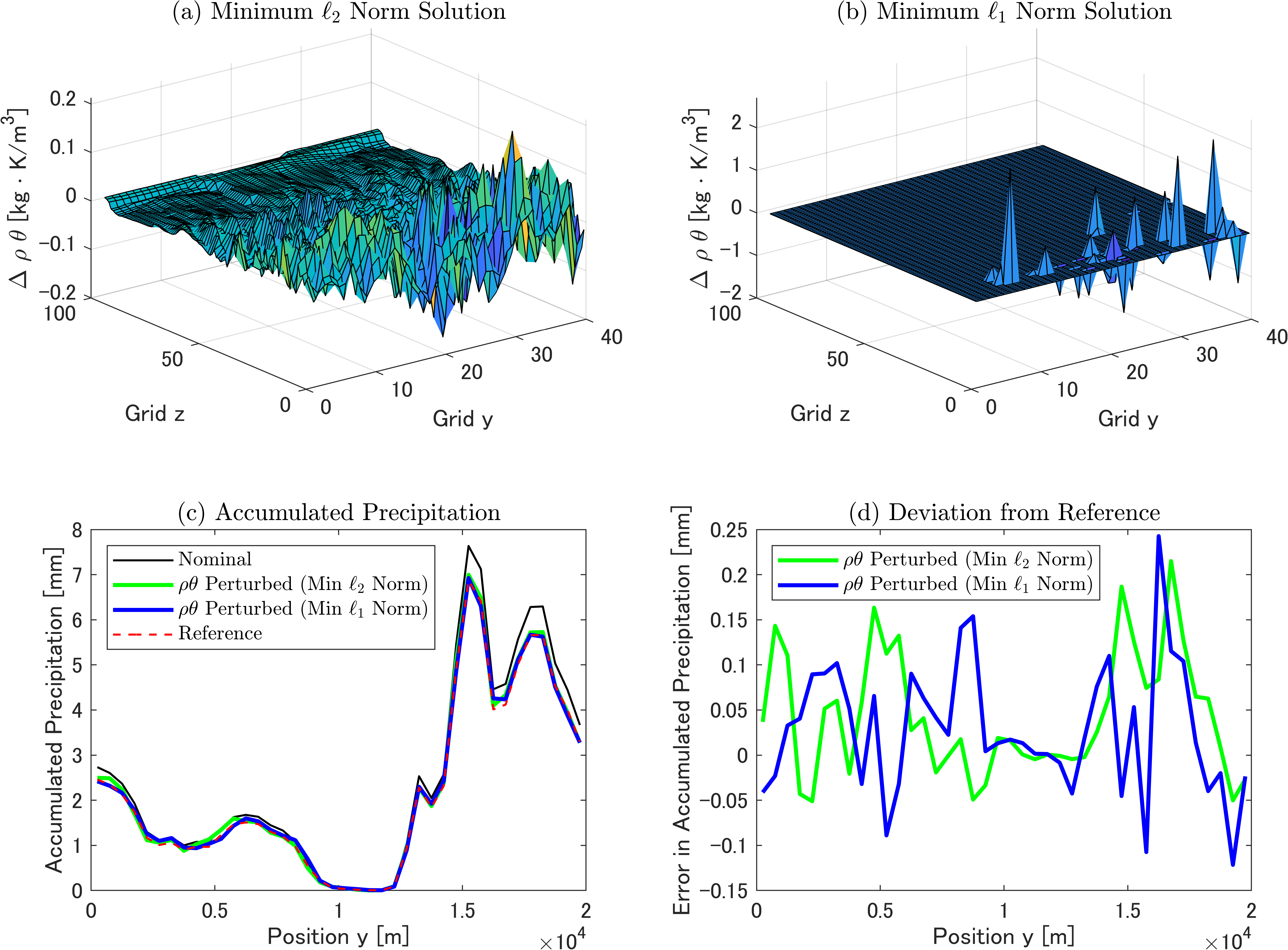} 
    \caption{(a) Minimum $\ell_2$ norm solution and (b) minimum $\ell_1$ norm solution of perturbations in the initial conditions of $\rho \theta$ for reducing accumulated precipitation to 90\% of the nominal case, (c) distributions of accumulated precipitation obtained by the NWP model for the nominal case (black solid line), perturbed cases (green and blue solid lines), and the reference (red dashed line), and (d) deviations of accumulated precipitation from the reference. }
    \label{fig:Ref_RHOT}
\end{figure}

Figures \ref{fig:Ref_RHOT}a and \ref{fig:Ref_RHOT}b show the minimum $\ell_2$ norm solution and the minimum $\ell_1$ norm solution of the perturbations in $\rho \theta$. 
Figure \ref{fig:Ref_RHOT}c shows distributions of accumulated precipitation for the nominal simulation result (black solid line), perturbed simulation results (green and blue solid lines) by SCALE-RM, and the reference value (red dashed line), respectively. 
Moreover, Fig.\ \ref{fig:Ref_RHOT}d shows deviations of accumulated precipitation from the reference in the perturbed simulation results. 
Figure S\ref{supp:perturbed} shows snapshots of perturbations in density $\rho$, potential temperature $\theta$, mass concentration of total hydrometeor, and precipitation intensity, respectively, for the minimum $\ell_1$ norm solution. 

In Fig.\ \ref{fig:Ref_RHOT}, the distribution of accumulated precipitation after perturbation is close to its reference distribution. This result justifies the use of the linear model and validates the proposed method to realize a desired spatial distribution of accumulated precipitation by small perturbation in the initial conditions. 
As the reference value is further reduced to 80\% (Fig.\ S\ref{supp:RHOT_Ref0p8}), the deviation of the perturbed simulation results from the reference values increases, indicating a slight deterioration in linearity. 
However, the initial perturbations still reduce the accumulated precipitation effectively. 

In both cases in Figs.\ \ref{fig:Ref_RHOT} and S\ref{supp:RHOT_Ref0p8}, the minimum $\ell_1$ norm solutions exhibit sparsity of the perturbations in the initial conditions compared to the minimum $\ell_2$ norm solutions. 
That is, we can realize the reference distribution of precipitation by perturbing the initial conditions at a limited number of grids determined by optimization. 
This sparsity is potentially advantageous in implementing weather control with small numbers of local interventions. 
On the other hand, the minimum $\ell_2$ norm solutions have smaller magnitudes and result in smaller maximum errors of the accumulated precipitation from the reference distributions than the minimum $\ell_1$ norm solutions.

\subsection{Reducing maximum precipitation}
Next, we considered a control problem of reducing the maximum value of accumulated precipitation to 90\% of that in the nominal case by perturbing the initial condition of $\rho \theta$ or $q_v$. 
This setting is potentially useful when the maximum value of accumulated precipitation determines the damage from heavy rain. 
In this case, we did not impose the equality constraint in Eq.\ (\ref{eq:CMNP}), and, instead, we imposed an inequality constraint $\vec{Y} + \mathbf{S}(\vec{X}^\mathrm{init}) \Delta \vec{X}^\mathrm{init} \leq (0.9 \max \vec{Y}) \vec{1}$, where $\max \vec{Y}$ denotes the maximum value among the components of nominal $\vec{Y}$, and $\vec{1}$ denotes a vector with all components one. 
This inequality constraint was recast as a linear inequality constraint $\mathbf{A} \Delta \vec{X}^\mathrm{init} \leq \vec{b}$ in Eq.\ (\ref{eq:GeneralCMNP}) with $\mathbf{A}=\mathbf{S}(\vec{X}^\mathrm{init})$ and $\vec{b} = (0.9 \max \vec{Y}) \vec{1} - \vec{Y}$. 
We also imposed the nonnegativity constraint $ -\Delta \vec{X}^\mathrm{init} \leq \vec{X}^\mathrm{init}$ for $q_v$. 
The computation times to solve the constrained norm minimization problems were 3 s and 4 s for $\rho \theta$ with the $\ell_2$ norm and $\ell_1$ norm, respectively, and 6 s and 4 s for $q_v$ with the $\ell_2$ norm and $\ell_1$ norm, respectively. 

\begin{figure*}[t]
    \centering
    \includegraphics[width=12cm]{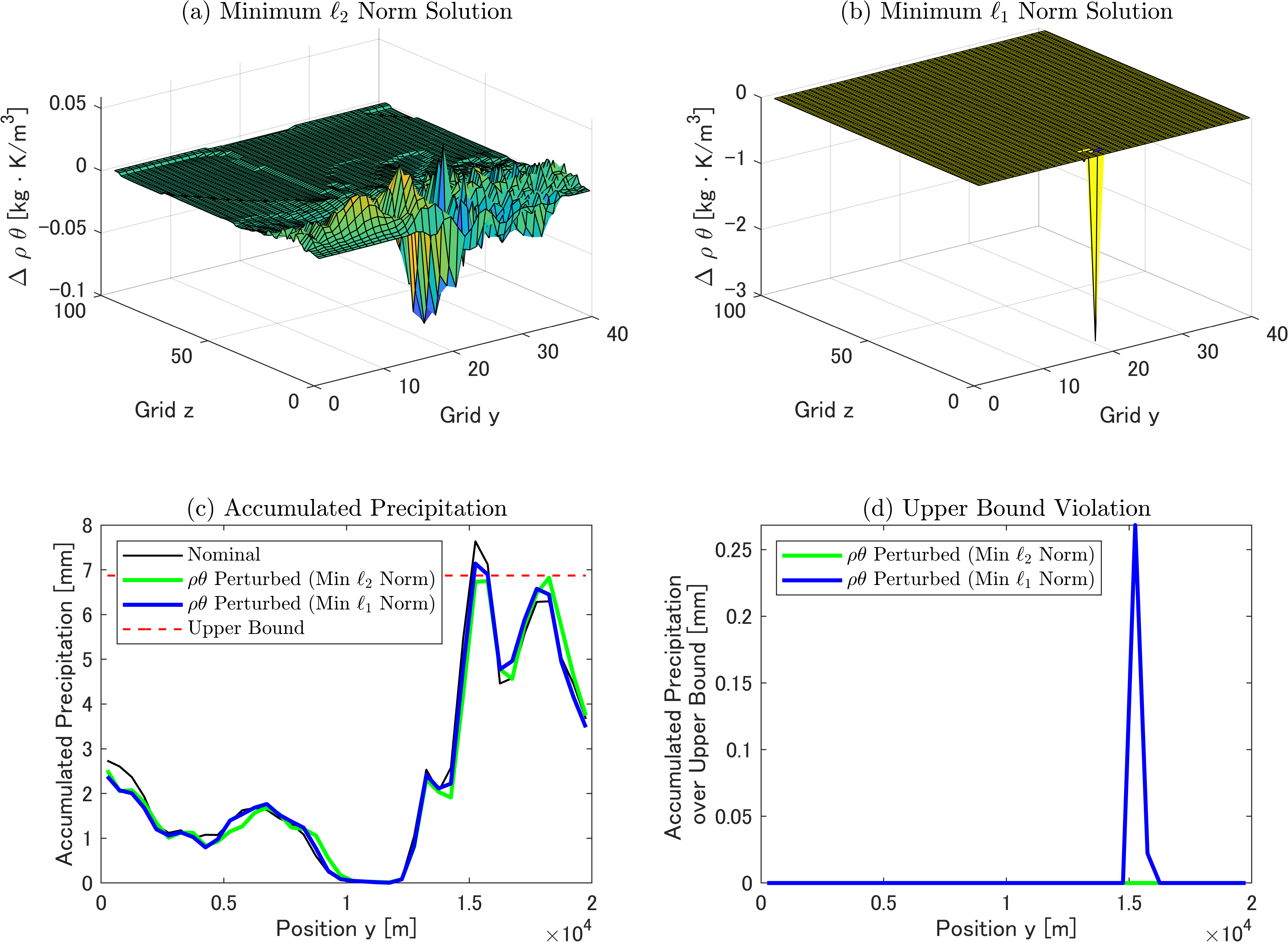}
    \caption{(a) Minimum $\ell_2$ norm solution and (b) minimum $\ell_1$ norm solution of perturbation in the initial conditions of $\rho \theta$ for upper-bounding accumulated precipitation to 90\% of the maximum value in the nominal case, (c) distributions of accumulated precipitation obtained by the NWP model for the nominal case (black solid line), perturbed cases (green and blue solid lines), and the upper bound (red dashed line), and (d) distributions of upper bound violation in accumulated precipitation. }
    \label{fig:Uub_RHOT}
\end{figure*}

\begin{figure*}[t]
    \centering
    \includegraphics[width=12cm]{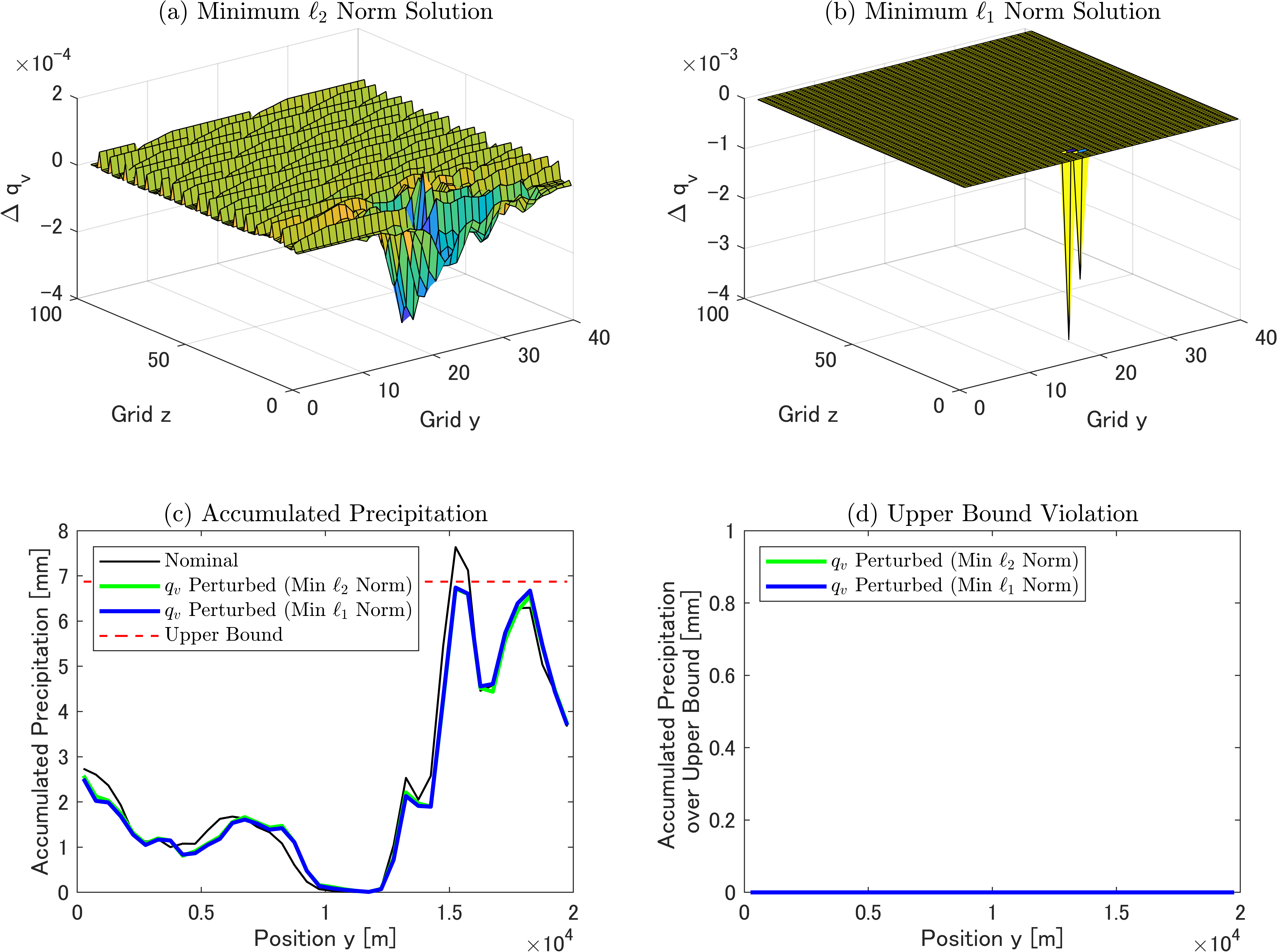}
    \caption{(a) Minimum $\ell_2$ norm solution and (b) minimum $\ell_1$ norm solution of perturbations in the initial conditions of $q_v$ for upper-bounding accumulated precipitation to 90\% of the maximum value in the nominal case, (c) distributions of accumulated precipitation obtained by the NWP model for the nominal case (black solid line), perturbed cases (green and blue solid lines), and the upper bound (red dashed line), and (d) distributions of upper bound violation in accumulated precipitation. }
    \label{fig:Uub_QV}
\end{figure*}

Figures \ref{fig:Uub_RHOT}a and \ref{fig:Uub_RHOT}b show the minimum $\ell_2$ norm solution and the minimum $\ell_1$ norm solution of the perturbations in $\rho \theta$. 
Figure \ref{fig:Uub_RHOT}c shows distributions of accumulated precipitation for the nominal simulation result (black solid line), perturbed simulation results (green and blue solid lines) by the nonlinear NWP model, SCALE-RM, and the upper bound (red dashed line), respectively. 
Moreover, Fig.\ \ref{fig:Uub_RHOT}d shows distributions of upper bound violation in accumulated precipitation in the perturbed simulation results. 
Then, Figs.\ \ref{fig:Uub_QV}a--\ref{fig:Uub_QV}d show the corresponding plots for the case where the initial values of $q_v$ are perturbed.  
In all cases, the accumulated precipitation at each grid after perturbation is bounded by the upper bound, except for the minor violation when $\rho \theta$ is perturbed by the minimum $\ell_1$ norm solution in Figs.\ \ref{fig:Uub_RHOT}c and \ref{fig:Uub_RHOT}d. 
The constraint violation is caused by a mismatch between the linear model for optimization and the highly nonlinear NWP model for validation. 
However, the constraint violation is minor, and the initial perturbations effectively reduce the accumulated precipitation. 
These results validate the proposed method to bound the spatial distribution of accumulated precipitation by perturbing the initial conditions. 
Moreover, the minimum $\ell_1$ norm solutions (Figs.\ \ref{fig:Uub_RHOT}b and \ref{fig:Uub_QV}b) of the perturbations in the initial conditions are more sparse than those for realizing reference precipitation in the previous Subsection \ref{subsec:RefPrec}. 
This higher sparsity is because the accumulated precipitation needs to be reduced only at fewer grid points. 

Figure S\ref{supp:RHOT_Uub0p8} shows corresponding plots where initial values of $\rho \theta$ are perturbed to reduce the maximum value of accumulated precipitation to 80\% of the nominal case. 
As the maximum value is further reduced, the imposed bound is violated at some points. 
However, the minimum $\ell_2$ norm solutions have smaller magnitudes resulting in smaller violations of the bound than the minimum $\ell_1$ norm solutions. 
In contrast, the minimum $\ell_1$ norm solutions have significantly smaller numbers of nonzero elements than the minimum $\ell_2$ norm solutions.

\section{Summary}

We proposed a convex optimization approach for quantitative weather control by perturbing the initial conditions as control inputs. 
We represented the control specifications as linear equality or inequality constraints consisting of the sensitivity matrix of an NWP model. 
Then we found optimal perturbations of initial conditions by minimizing their norms under the constraints. 
The constrained norm minimization problems in the numerical experiments were solved in a matter of seconds for thousands of variables, indicating the high numerical efficiency of the proposed approach. 
The numerical experiments also showed control specifications on spatial distributions of accumulated precipitation were realized in an NWP model. 
In particular, the number of grids to perturb the initial conditions was significantly small when minimizing the $\ell_1$ norm, which is potentially advantageous in implementing weather control with small numbers of local interventions. 
Based on these results, the proposed method is promising and provides the basis for developing techniques of quantitative weather control. 


%
\section*{Acknowledgments}

The authors thank Kazuaki Yasunaga and Yusuke Hiraga for valuable discussions about the physical processes of the warm bubble experiment. 
This work was supported by JST Moonshot R\&D Program Grant Numbers JPMJMS2284 and JPMJMS2389-1-1, 1-2, 4-1 and 4-2. 

%

\begin{supplements}
\renewcommand{\labelenumi}{Figure S\theenumi}
\item \label{supp:nominal} Snapshots of mass concentration of (a)--(f) total hydrometeor (QHYD) and (g)--(l) precipitation intensity (PREC) in warm bubble experiment.  

\item \label{supp:perturbed} Snapshots of perturbations in (a)--(f) density $\rho$, (g)--(l) potential temperature $\theta$, (m)--(r) mass concentration of total hydrometeor (QHYD), and (s)--(x) precipitation intensity (PREC) with initial conditions of $\rho \theta$ perturbed according to the minimum $\ell_1$ norm solution for reducing accumulated precipitation to 90\% of the nominal case. At $t = 1500$ [s] and $t = 2100$ [s], an increase in density $\rho$ and a decrease in potential temperature $\theta$ are observed at the top of the warm bubble. Such changes indicate that the perturbation suppressed the growth of the warm bubble and reduced precipitation. 

\item \label{supp:RHOT_Ref0p8} (a) Minimum $\ell_2$ norm solution and (b) minimum $\ell_1$ norm solution of perturbations in the initial conditions of $\rho \theta$ for reducing accumulated precipitation to 80\% of the nominal case, (c) distributions of accumulated precipitation obtained by the NWP model for the nominal case (black solid line), perturbed cases (green and blue solid lines), and the reference (red dashed line), and (d) deviations of accumulated precipitation from the reference. 

\item \label{supp:RHOT_Uub0p8} (a) Minimum $\ell_2$ norm solution and (b) minimum $\ell_1$ norm solution of perturbations in the initial conditions of $\rho \theta$ for upper-bounding accumulated precipitation to 80\% of the maximum value in the nominal case, (c) distributions of accumulated precipitation obtained by the NWP model for the nominal case (black solid line), perturbed cases (green and blue solid lines), and the upper bound (red dashed line), and (d) distributions of upper bound violation in accumulated precipitation. 
\end{supplements}

%

\bibliographystyle{sola}
\bibliography{SOLAref}

%

%


\end{document}